\documentclass[%
 reprint,
superscriptaddress,
 amsmath,amssymb,
prl,
]{revtex4-2}

\usepackage{graphicx}
\usepackage{dcolumn}
\usepackage{bm}
\usepackage{color}
\usepackage{xspace}
\usepackage{multirow}
\usepackage{hyperref}
\usepackage{ulem}
\usepackage{here}

\definecolor{green}{rgb}{0,0.6,0.1}

\begin{document}

\preprint{APS/123-QED}

\title{Fermi surface expansion above critical temperature in a Hund ferromagnet}

\author{Yusuke Nomura}
\email{yusuke.nomura@riken.jp}
\affiliation{
RIKEN Center for Emergent Matter Science, 2-1 Hirosawa, Wako, Saitama 351-0198, Japan
}

\author{Shiro Sakai}
\affiliation{
RIKEN Center for Emergent Matter Science, 2-1 Hirosawa, Wako, Saitama 351-0198, Japan
}

\author{Ryotaro Arita}
\affiliation{
RIKEN Center for Emergent Matter Science, 2-1 Hirosawa, Wako, Saitama 351-0198, Japan
}
\affiliation{
Department of Applied Physics, The University of Tokyo,7-3-1 Hongo, Bunkyo-ku, Tokyo 113-8656
}

\date{\today}

\begin{abstract}
Using a cluster extension of the dynamical mean-field theory, 
we show that strongly correlated metals subject to Hund's physics exhibit significant electronic structure modulations above magnetic transition temperatures.
In particular, in a ferromagnet having a large local moment due to Hund's coupling (Hund's ferromagnet), the Fermi surface expands even above the Curie temperature ($T_{\rm C}$) as if a spin polarization occurred. 
Behind this phenomenon, effective ``Hund's physics'' works in momentum space, originating from ferromagnetic fluctuations in the strong coupling regime.
The resulting significantly momentum-dependent (spatially nonlocal) electron correlations induce an electronic structure reconstruction involving a Fermi-surface volume change and a redistribution of the momentum-space occupation.  
Our finding will give a deeper insight into the physics of Hund's ferromagnets above $T_{\rm C}$.
\end{abstract}

\maketitle


\paragraph{Introduction.}

The concept of the Fermi liquid offers a firm basis to understand the behavior of interacting electrons in metals. 
However, properties of strongly correlated metals often deviate from the Fermi-liquid behavior.
A famous example is a pseudogap or strange metal behavior seen in doped cuprates~\cite{Keimer_2015}. 
Since various fascinating phenomena such as superconductivity and magnetism emerge from these states, understanding correlation effects beyond the Fermi-liquid theory poses a fundamental issue in condensed matter physics. 
Although the origin of the pseudogap in the cuprates is still controversial, one of the widely discussed scenarios is an antiferromagnetic(AFM)-fluctuation-induced mechanism~\cite{Kampf_1990,Millis_1990,Moriya_1990,Gunnarsson_2015}: Even without symmetry breaking, the AFM fluctuation induces a significant momentum-dependent self-energy, giving rise to the pseudogap-like behavior.

Recently, a ferromagnetic (FM) quantum critical point and associated strange metal behavior were observed in a heavy-fermion material CeRh$_6$Ge$_4$~\cite{Shen_2020}.  
The effect of a FM quantum critical point was also studied for a tailored Hamiltonian in which itinerant fermions couple to a critical FM bosonic mode of the transverse-field Ising model, and non-Fermi liquid behavior was reported~\cite{Xu_2017,Xu_2020}. 
Unusual spectral properties involving a splitting of the spectrum in the vicinity of a FM instability were reported in Refs.~\cite{Katanin_2005,Katanin_2008}.

However, in comparison to the metallic state subject to AFM fluctuations, our understanding of that under FM fluctuations is still scarce. 
Of particular interest is a strong-coupling regime, where the local moment is formed well above the Curie temperature ($T_{\rm C}$) and affects the metallic behavior. 
This may result in a non-Fermi liquid state distinct from the strong-coupling AFM one, where the Mott physics renders the local-moment formation above the N\'eel temperature.

Such a strong-coupling FM regime is expected in the multiorbital Hubbard model, where Hund's coupling stabilizes a large local moment.
This large local moment indeed gives rise to strong {\it local} correlations~\cite{Werner_2008,Haule_2009,Georges_2013}, as revealed by the previous studies based on the dynamical mean-field theory (DMFT)~\cite{Metzner_1989,Georges_1996}. 
Beyond the DMFT, Hund's coupling also induces significant {\it nonlocal} correlations~\cite{Nomura_2015} while its effect on spectral properties remains unexplored.

In this Letter, we study the two-orbital Hubbard model with the cellular dynamical mean-field theory  (cDMFT)~\cite{Lichtenstein_2000,Kotliar_2001}, a cluster extension~\cite{Maier_2005} of the DMFT, revealing how the Hund-induced nonlocal spin correlations influence the single-particle properties.
The model exhibits two distinct regimes, one governed by the AFM fluctuation and the other governed by the FM fluctuation, depending on the electron filling.
We first show that the different types of spin correlations bring about different momentum dependencies of the electron self-energy, reconstructing the Fermi surface (FS) differently.

Then, focusing on the FM-fluctuation regime, we find a FS-volume expansion {\it above} $T_{\rm C}$: While it is natural to see a volume expansion of the majority-spin FS in the FM long-range-ordered metal below $T_{\rm C}$, we find a similar spectrum even above $T_{\rm C}$ in the strong-coupling regime with the large preformed local moment.
We discuss the emergence of ``Hund's physics'' in momentum space behind this unusual behavior.

\paragraph{Model and Methods.}

We study a degenerate two-orbital Hubbard model on a square lattice. 
The interaction part of the Hamiltonian reads
\begin{eqnarray}
 \mathcal{H}_{\rm int} 
&=&   
 U \sum_{i,l} n^{\uparrow}_{l i}  n^{\downarrow}_{ l i } 
+   
U'    \sum_{i,\sigma} 
 n^{\sigma}_{1 i}  n^{\! -\sigma} _{2 i}   +
(U' \! -\! J)  \sum_{i,\sigma}  
n^{\sigma}_{1 i}  n^{\sigma}_{2 i} 
 \nonumber \\ 
 &+& 
 J \!  \sum_{i, l \neq m }  
c^{\uparrow \dagger}_{li} 
c^{\uparrow}_{mi} 
c^{\downarrow \dagger}_{mi}
c^{\downarrow}_{li}
+   
J  \!  \sum_{i, l \neq m} 
c^{\uparrow \dagger }_{li} 
c^{\uparrow}_{mi} 
c^{\downarrow \dagger}_{li}
c^{\downarrow}_{mi}, 
\label{hamiltonian}
\end{eqnarray}
where $c_{l i}^{\sigma \dagger}$ ($c^{\sigma}_{l i}$) creates (annihilates) an $l$th-orbital electron ($l=1,2$) with spin $\sigma$ at site $i$ and $n^{\sigma}_{l i}\equiv \hat{c}_{l i}^{\sigma \dagger} \hat{c}^{\sigma}_{l i}$.
The intra- and inter-orbital Coulomb interactions ($U$ and $U'$), and Hund's coupling ($J$) satisfy $U'=U-2J$ to keep the Hamiltonian rotationally invariant in orbital and spin spaces. 
We consider the noninteracting dispersion of $\varepsilon_{{\bf k}}=-2t(\cos k_x + \cos k_y)-4t'\cos k_x \cos k_y$ for each orbital [$t$ and $t'$ are nearest-neighbor (NN) and next-nearest-neighbor (NNN) hoppings, respectively]. We set $t=1$ as the energy unit and study the case of $t'/t = -0.2$ and $J/U = 1/4$.
We consider the paramagnetic and paraorbital phase.
Then, the Green's function is given by $G_{\bf k}( i \omega_\nu )  = \bigl [ i \omega_\nu + \mu - \varepsilon_{\bf k}  - \Sigma_{\bf k} (i \omega_\nu ) \bigr]^{-1}$ with the Matsubara frequency $\omega_\nu = (2\nu + 1 ) \pi T$ ($T=1/\beta$ is the temperature), the chemical potential $\mu$, and the self-energy $\Sigma_{\bf k} (i \omega_\nu )$. 
Hereafter, we omit spin and orbital indices for simplicity.

Within the cDMFT, we map the lattice model onto a $2\times 2$ cluster impurity problem under a self-consistent condition. 
Solving the latter model numerically, we incorporate short-range correlations within the cluster size. 
We use a continuous-time quantum Monte Carlo method with an interaction expansion~\cite{Rubtsov_2004,Rubtsov_2005} developed in Ref.~\cite{Nomura_2014} to solve the impurity problem.
It is computationally demanding to solve a cluster multi-orbital impurity problem because of the large expansion order ($\sim 1000$) and the sign problem. 
We have implemented the submatrix update~\cite{Gull_2011} to handle the large expansion order and the double-vertex update~\cite{Nomura_2014} to mitigate the sign problem. 
Even with these techniques, at a temperature of $T \lesssim W/100$ ($W$: bandwidth), the average sign becomes less than 0.01 in the worst case. 
When considering the expansion order and the sign problem, the $2\times 2$ cluster is the maximum cluster size that we can handle within a realistic computational time
 \footnote{An open-source CT-INT package using the submatrix update as in Ref.~\cite{Nomura_2014} is now available in Ref.~\cite{Shinaoka_2020}.}.

\begin{figure}[tb]
\vspace{0.0cm}
\begin{center}
\includegraphics[width=0.48\textwidth]{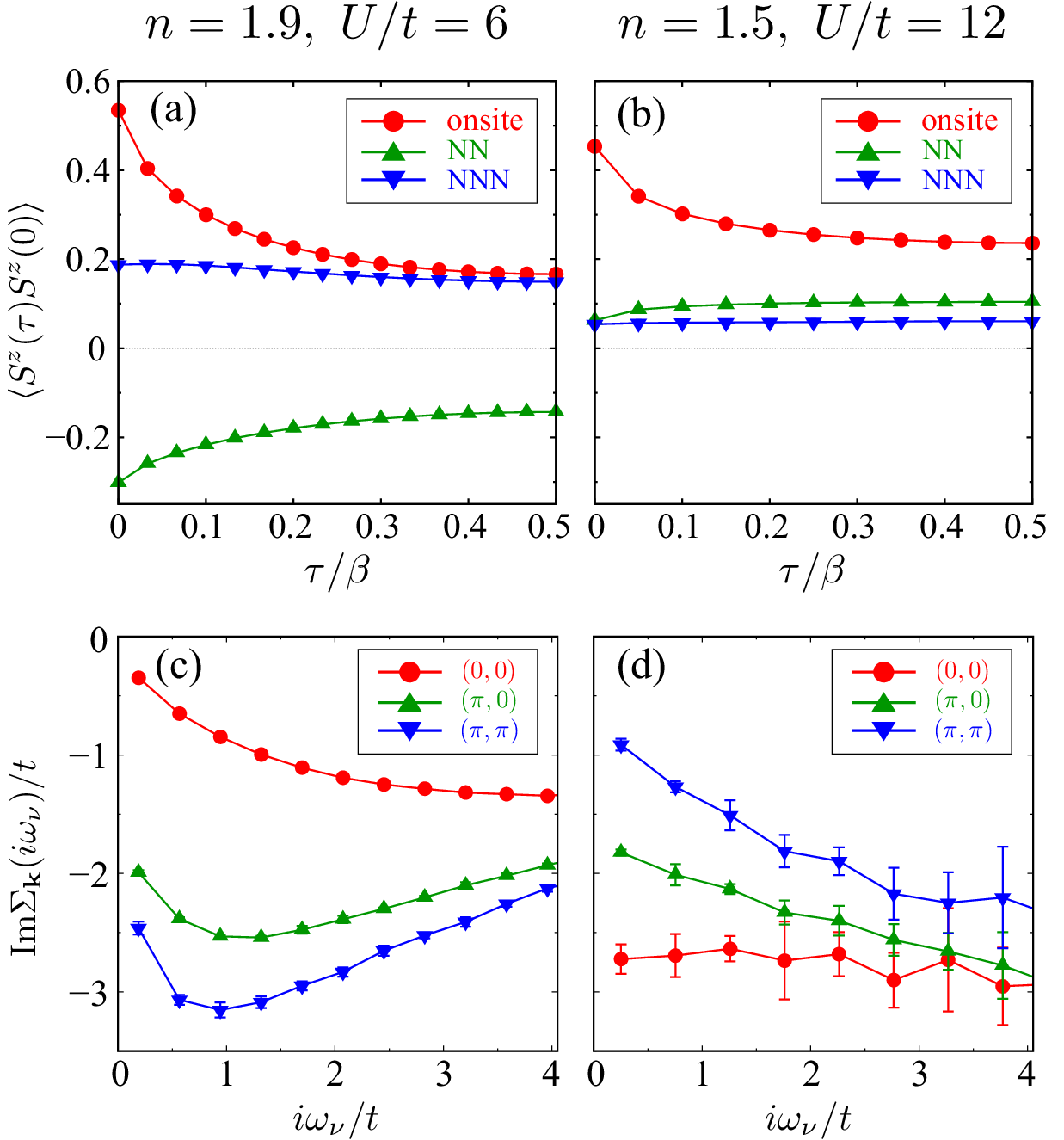}
\caption{
Real-space spin-spin correlation function and imaginary part of the self-energy at the momenta $(0,0)$, $(\pi,0)$ [or equivalently $(0,\pi)$], and $(\pi,\pi)$. 
(a,c) Results for $n=1.9$ and $U=4J=6$. 
(b,d) Results for $n=1.5$ and $U=4J=12$.
}
\vspace{-0.2cm}
\label{Fig_chi_s_sigma_raw}
\end{center}
\end{figure}

\paragraph{Results.}
We study two different fillings $n=1.9$ (5 \% hole doping from the half filling) and $n=1.5$ (25 \% hole doping).
The temperature is set to $T=0.06$ ($0.08$\footnote{The temperature of $T=0.08$ is above the FM critical temperature estimated by the single-site DMFT for the present model ($T_{\rm C} \approx0.05$-0.06).}) for $n=1.9$ ($1.5$). 
The former filling favors the AFM correlation [Fig.~\ref{Fig_chi_s_sigma_raw}(a)], whereas the latter favors the FM correlation [Fig.~\ref{Fig_chi_s_sigma_raw}(b)] at large interaction parameters due to the double exchange mechanism.
The appearance of these magnetic correlations is qualitatively similar to the single-site DMFT results~\cite{Held_1998,Momoi_1998,Sakai_2007,Hoshino_2016}.
Under strong Hund's coupling, the paramagnetic solution of the DMFT shows a spin-freezing behavior~\cite{Werner_2008}, which is also seen in Figs.~\ref{Fig_chi_s_sigma_raw}(a) and \ref{Fig_chi_s_sigma_raw}(b) for the onsite correlation function:
The slow decay against the imaginary time $\tau$ implies a frozen ($\omega\!=\!0$) component.
In addition, our cDMFT results show a freezing behavior also in the nonlocal (NN and NNN) spin-spin correlations.
As the local unscreened frozen moment induces local electron correlations giving bad metallic behavior~\cite{Werner_2008}, the frozen nonlocal spin correlation might induce exotic nonlocal electron correlations, which are the main subject of this study.

Indeed, at both fillings, the self-energy shows bad metallic behavior with a strong momentum dependence [Figs.~\ref{Fig_chi_s_sigma_raw}(c,d)]. 
Interestingly, however, the momentum dependence differs qualitatively between the two fillings.

First, for $n=1.9$,
due to the strong nonlocal AFM correlation, the self-energy becomes large around ${\bf k}=(\pi,\pi)$ [Fig.~\ref{Fig_chi_s_sigma_raw}(c)].
Although the 2$\times$2 cDMFT gives a coarse momentum resolution of $(0,0)$, $(\pi,0)$, $(0,\pi)$ and $(\pi,\pi)$, the fully momentum-dependent self-energy can be inferred through the periodization scheme~\cite{Kotliar_2001}. 
Here, we use the cumulant periodization~\cite{Stanescu_2006}, which gives a fast convergence of the periodized self-energy against the cluster size: 
For the single-orbital model, 
the results of the $2 \times 2 $ cluster qualitatively agree with the converged results at a temperature similar to that used in this study~\cite{Sakai_2012}.

\begin{figure}[b]
\vspace{0.0cm}
\begin{center}
\includegraphics[width=0.48\textwidth]{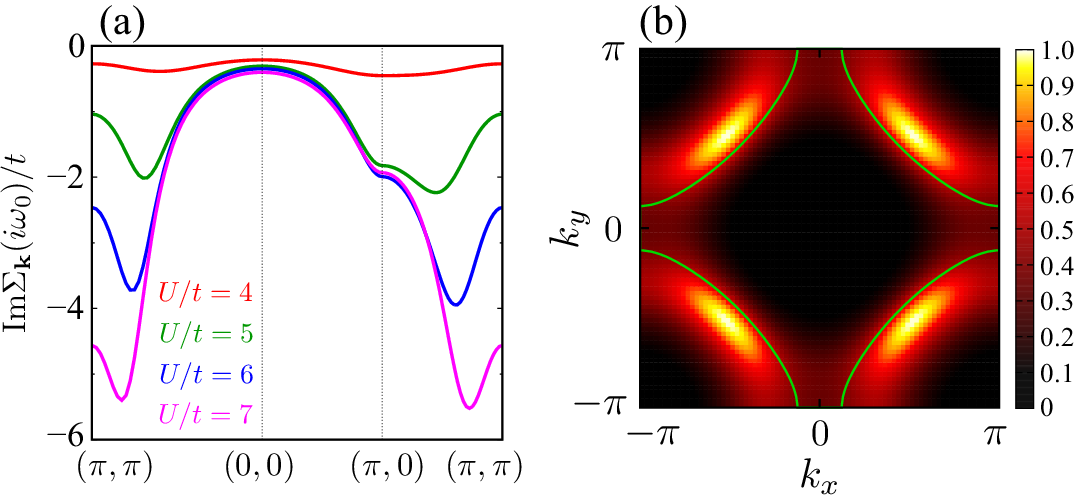}
\caption{
Results for $n=1.9$.  
(a) Imaginary part of the periodized self-energy for various values of $U=4J$. (b) The FS [$-\beta G_{\bf k} (\tau \! = \! \beta/2)$ normalized by its maximum value] at $U=4J=6$.
The light green curve in (b) shows the FS at $U=J=0$. 
}
\vspace{-0.5cm}
\label{Fig_n1.9_FS_Sigma}
\end{center}
\end{figure}

\begin{figure*}[tb]
\vspace{0.0cm}
\begin{center}
\includegraphics[width=0.99\textwidth]{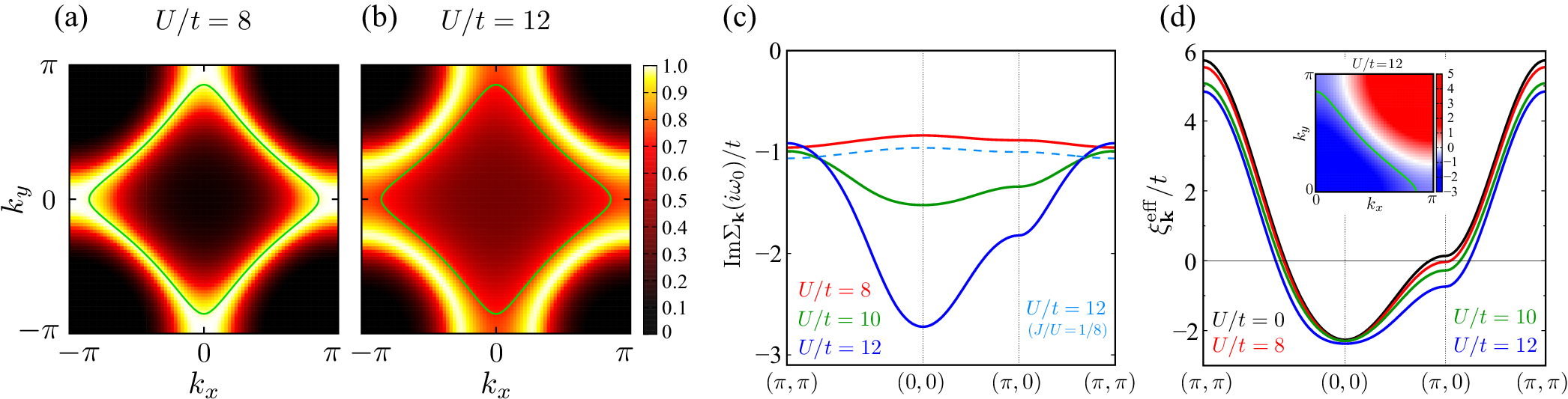}
\caption{
Results for $n=1.5$ and $J/U=1/4$.  
(a,b) The FS [$-\beta G_{\bf k} (\tau \! = \! \beta/2)$ normalized by its maximum value], 
(c) the imaginary part of the periodized self-energy, and
(d) the effective dispersion $\xi_{\bf k}^{\rm eff} = \epsilon_{\bf k} + {\rm Re }\Sigma_{\bf k}(0) - \mu $ [where ${\rm Re }\Sigma_{\bf k}(0)$ is approximated by ${\rm Re }\Sigma_{\bf k}( i \omega_0)$] with the inset showing the contour map of $\xi_{\bf k}^{\rm eff}$ for $U=12$. 
The light green curves in (a,b) 
and the inset of (d)
show the FS at $U=J=0$. 
In (c), for comparison, the result for $U=8J=12$ (dashed curve) is also shown. 
}
\vspace{-0.3cm}
\label{Fig_n1.5_FS_Sigma}
\end{center}
\end{figure*}

The periodized self-energy is shown in Fig.~\ref{Fig_n1.9_FS_Sigma}(a).
It shows qualitatively similar behavior to that of the single-orbital model [Fig.~\ref{Fig_1orb_AFM}(a) in supplemental materials (SM)\footnote{See Supplemental Materials at http://***.}] at the same doping ratio (5 \%). 
However, in the two-orbital model, strong momentum dependence develops at smaller interaction values than those in the single-orbital model.  
This confirms a significant role of Hund's coupling in inducing strong electron correlations through forming a large local moment.

As a result of the strongly momentum-dependent self-energy, the FS is considerably modified. Figure~\ref{Fig_n1.9_FS_Sigma}(b) shows this by approximating the spectral weight at the Fermi level by $-\beta G_{\bf k} (\tau \! = \! \beta/2)$.
The FS shows hot and cold spots, and the scattering rate shows a strong angle dependence. 
This feature is qualitatively similar to that of the single-orbital model [Fig.~\ref{Fig_1orb_AFM}(b) in SM], 
as well as to angle-resolved photoemission spectroscopy results for the cuprates~\cite{Norman_1998,Keimer_2015}.
The results suggest that an exotic pseudogap-like feature might also be seen in the multiorbital systems with strong AFM correlations.

Next, we turn to the 25 \% hole-doping ($n=1.5$) case.
In the single-orbital model, the filling corresponds to the overdoped regime of cuprates, where the Fermi-liquid behavior is observed. 
However, in the two-orbital model, we find a strong FM spin correlation due to Hund's coupling, where the local moment is formed well above $T_{\rm C}$.
This poses an intriguing issue: 
How does the strong FM correlation affect the single-particle quantities?
As we see in Fig.~\ref{Fig_chi_s_sigma_raw}(d), in stark contrast with the AFM case, the self-energy becomes largest at around ${\bf k}=(0,0)$. 
Then, the FS shape will be modulated differently.

Using the periodization method again, we investigate the change of the FS between $U=8$ and 12 [Figs.~\ref{Fig_n1.5_FS_Sigma}(a) and \ref{Fig_n1.5_FS_Sigma}(b)]. 
At $U=8$, the momentum dependence of the self-energy is not significant [red curve in Fig.~\ref{Fig_n1.5_FS_Sigma}(c)].
The nonlocal spin correlation is not large, either [Fig.~\ref{Fig_n1.5_chi_s}(a) in SM]. 
Nevertheless, the spin freezing behavior is seen for the local spin correlation. 
This makes the quasiparticle lifetime short and broadens the low-energy spectral weight [Fig.~\ref{Fig_n1.5_FS_Sigma}(a)]. 
These features are consistent with the Hund's metal behavior investigated intensively with the single-site DMFT~\cite{Georges_2013}.

However, as $U$ increases, the nonlocal FM correlation grows. 
Simultaneously, the self-energy acquires a significant momentum dependence [Fig.~\ref{Fig_n1.5_FS_Sigma}(c)].
Figure~\ref{Fig_n1.5_FS_Sigma}(b) shows the FS with the strongly momentum-dependent self-energy at $U=12$.
In contrast to the $n=1.9$ case, we do not see a clear angle dependence of the scattering rate on the FS. 
However, interestingly, the volume inside the FS expands compared to the noninteracting case.  
In the Fermi liquid, the volume should not change according to Luttinger's theorem~\cite{Luttinger_1960}. 
Thus, a change in the volume indicates an appearance of an unusual metallic state. 
Note that, for a smaller Hund's coupling ($J/U=1/8$), the momentum dependence of the self-energy becomes much weaker [dashed curve in Fig.~\ref{Fig_n1.5_FS_Sigma}(c)]. 
This evidences that Hund's coupling is the trigger of the unusual nonlocal correlation effect.

To further analyze the FS at $n=1.5$, in Fig.~\ref{Fig_n1.5_FS_Sigma}(d), we show the effective single-particle energy dispersion $\xi_{\bf k}^{\rm eff} = \epsilon_{\bf k} + {\rm Re }\Sigma_{\bf k}(0) - \mu $, where ${\rm Re }\Sigma_{\bf k}(0)$ is approximated by ${\rm Re }\Sigma_{\bf k}( i \omega_0)$. 
$\xi_{\bf k}^{\rm eff}=0$ determines the position of the FS, unless the imaginary part of the self-energy is large.
As the interaction increases, the effective dispersion is modified by the correlation effect. 
As is clear from the inset of Fig.~\ref{Fig_n1.5_FS_Sigma}(d),
the momenta for $\xi_{\bf k}^{\rm eff}=0$ at $U=12$ (white region) deviate from the noninteracting ones (green curve),
consistent with the FS expansion shown in Fig.~\ref{Fig_n1.5_FS_Sigma}(b). 
We note that at the momenta ${\bf k}=(\pi,0)$ and $(0,\pi)$, directly accessible by the $2\times2$ cDMFT, $\xi_{\bf k}^{\rm eff}$ is negative for $U=12$ (i.e., ${\bf k}=(\pi,0)$ and $(0,\pi)$ are inside the FS), in contrast with the $U=0$ case showing a positive $\xi_{\bf k}^{\rm eff}$
\footnote{The sign change in $\xi_{\bf k}^{\rm eff}$ value at ${\bf k} = (\pi,0)$ from $U=0$ to $U=12$ is robust against the way of estimating $\xi_{{\bf k}}^{\rm eff}$.
 If we approximate ${\rm Re }\Sigma_{\bf k}(0)$ by ${\rm Re }\Sigma_{\bf k}( i \omega_0)$ as in the main text, the $\xi_{(\pi,0)}^{\rm eff}$ value is $\xi_{(\pi,0)}^{\rm eff}=-0.741(5)$ at $U=12$.
Even when we estimate ${\rm Re }\Sigma_{\bf k}(0)$ by a linear extrapolation of ${\rm Re }\Sigma_{\bf k}( i \omega_0)$ and ${\rm Re }\Sigma_{\bf k}( i \omega_1)$, we obtain $\xi_{(\pi,0)}^{\rm eff}=-0.645(8)$.}.
Thus, the volume expansion of the FS is not an artifact of the periodization.

Under the strong freezing FM correlations, even though the long-time average is zero, a spin-polarized state is realized as a snapshot. 
Then, it acts like an effective ``Hund's coupling'' in momentum space, 
as the FM exchange interaction in real space aligns spins also in momentum space~\cite{Nomura_2015}.
Then, the effective ``Hund's coupling'' competes with an effective ``crystal-field splitting'' (energy difference in $\epsilon_{\bf k}$ between different momenta). 
As Hund's coupling effectively reduces the crystal-field splitting and pump electrons into unoccupied orbitals to realize high-spin states in real space, the effective ``Hund's coupling'' brings about the rearrangement of the electron distribution in momentum space.

\begin{figure}[tb]
\vspace{0.0cm}
\begin{center}
\includegraphics[width=0.48\textwidth]{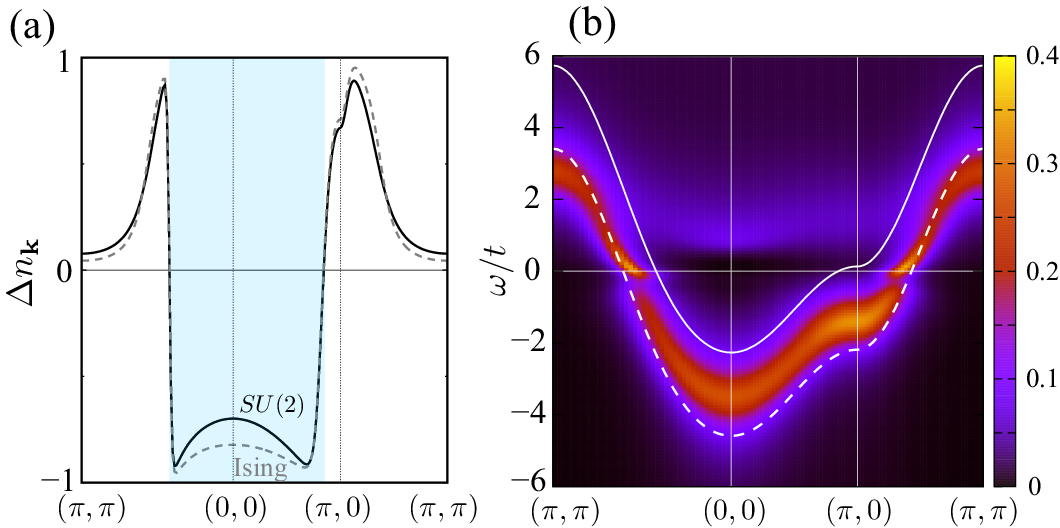}
\caption{
(a) Change in the momentum-space occupation $\Delta n_{\bf k}$ per orbital from the noninteracting case, calculated for $n=1.5$, $U=4J=12$ and $T=0.08$.
The shaded part indicates the region inside the FS at $U=0$.
The result with the density-density-type interaction (dotted curve) is also shown for comparison.  
(b) Spectral function $A_{\bf k}(\omega) = - \frac{1}{\pi} {\rm Im } G_{\bf k} (\omega) $ at $n=1.5$, $U=4J=12$, and $T=0.06$ for the density-density-type interaction model, where the correlation effect is exaggerated compared to the rotationally invariant case.
The solid (dashed) curve indicates the noninteracting (fully polarized ferromagnetic majority-spin) band dispersion. 
For the analytic continuation, we used the {\sc Maxent} package~\cite{Levy_2017}.
}
\vspace{-0.3cm}
\label{Fig_Delta_Nk_Akw}
\end{center}
\end{figure}

The effective ``Hund's coupling'' not only affects the FS but also rearranges the electron occupation $n_{\bf k}$ in the whole momentum space. 
Figure~\ref{Fig_Delta_Nk_Akw}(a) shows the change $\Delta n_{\bf k}$ of the occupation from the noninteracting value. 
Inside the original FS at $U=0$, $\Delta n_{\bf k}$ is close to $-1$, which means that $n_{\bf k}$ is close to half-filling. 
This is due to the strong imaginary part of the self-energy around ${\bf k}=(0,0)$ [Fig.~\ref{Fig_n1.5_FS_Sigma}(c)], which reduces the occupation significantly inside the original FS by producing the incoherent weight on the unoccupied side~\footnote{A similar momentum-space redistribution is observed in the two-site DCA (dynamical cluster approximation) study for the three-orbital model in Ref.~\cite{Nomura_2015}.}.
To compensate the occupation loss around ${\bf k}=(0,0)$, 
the occupation increases significantly outside the original FS and inside the expanded FS: in this region, the effective ``Hund's coupling'' lowers the energy of the originally unoccupied momenta, making $\xi_{\bf k}^{\rm eff}$ negative [Fig.~\ref{Fig_n1.5_FS_Sigma}(d)].

When the interaction is restricted to the density-density type,
i.e., Hund's coupling is assumed to be Ising-like with only $S_zS_z$ components, the correlation effect is exaggerated compared to the original rotationally invariant case.
The self-energy at ${\bf k} = (0,0)$ becomes more divergent, and the reconstruction of the FS is seen more clearly (Figs.~\ref{Fig_Ising_n1.5_Sigma_00} and \ref{Fig_Ising_n1.5_T0.06_FS} in SM). 
Also, the momentum-space redistribution of the electron occupation is more significant [Fig.~\ref{Fig_Delta_Nk_Akw}(a)]: The occupation inside the original FS is closer to half-filling $n_{\bf k}=1$. 
This shows more clearly that, in the unusual metallic state, the effective ``Hund's coupling'' surpasses the effective ``crystal-field splitting'', forming a ``high-spin''-like configuration in momentum space.

Figure~\ref{Fig_Delta_Nk_Akw}(b) shows the single-particle spectral function in the strong-coupling FM-fluctuation regime 
($U=12$, $T=0.06$ with the density-density-type interaction). 
The peak of the spectral function 
deviates significantly (on the order of $t$)
from the noninteracting paramagnetic band dispersion (solid curve). 
Rather, it is close to the majority-spin band dispersion (dashed curve) with a mass renormalization.
We also see a feature around $\omega=1$ that looks like a blurred minority-spin band.  
Thus, even in the paramagnetic state, we see a spectral feature as if a spin polarization occurred\footnote{A Fermi surface modulation under long-range FM correlations was discussed in Refs.~\cite{Katanin_2005,Katanin_2008}.
In contrast, we study the effect of short-range FM correlations in the strong-coupling regime.}. 
As we shall discuss below, this is characteristic of the strong-coupling regime, where the local moment is formed well above $T_{\rm C}$.

\paragraph{Discussion.}
Recently, unusual metallic behaviors have been reported under strong FM fluctuations~\cite{Shen_2020,Xu_2017,Xu_2020} as is mentioned in the introduction. 
Whereas these studies consider the coupling of itinerant fermions to critical spin systems,  
our study suggests that the $d$-electron systems represented by the multiorbital Hubbard model may offer another excellent playground to study 
unusual metallic behaviors subject to FM fluctuations.
Although there exists a FM fluctuating regime~\cite{Honerkamp_2001,Katanin_2003} in the single-orbital Hubbard model with a large $t'$ around 50 \% hole doping, 
the correlation effect is rather weak due to the small electron density (see Fig.~\ref{Fig_1orb_FM} in SM).
This suggests that Hund's ferromagnets in the strong coupling regime, where unscreened large local moments are fluctuating above $T_{\rm C}$, are suitable to see the 
FS expansion
clearly.


The present cDMFT results go beyond the conventional Hund's metal picture discussed within the DMFT. 
The strongly momentum-dependent self-energy modifies the effective dispersion and brings about the momentum-space electron redistribution through producing the incoherent spectral weight. 
Furthermore, in contrast with the DMFT result, in which a Fermi-liquid behavior recovers at very low temperatures~\cite{Georges_2013,Stadler_2015},
the divergence of the self-energy at ${\bf k}=(0,0)$ becomes significant as $T$ decreases (Fig.~\ref{Fig_Ising_n1.5_Sigma_00} in SM).
Thus, the present bad metallic behavior at finite temperatures may persist down to low temperatures, and a non-Fermi liquid phase, which cannot be connected adiabatically to the Fermi-liquid fixed point, may emerge at $T=0$.

\paragraph{Summary.}

Using the cDMFT, we have revealed a different aspect of Hund's physics discussed so far locally in real space. 
Associated with the strong AFM and FM fluctuations, the electron self-energy acquires a strong momentum dependence in different ways. 
In particular, the FM spin correlation induces effective ``Hund's physics'' in momentum space, leading to a significant modulation of the momentum-space occupation. 
The resulting FS shows a volume change at finite temperatures, differently from the AFM-fluctuating metals.
Finally, we note that, to detect the unusual metallic state experimentally, FM materials exhibiting large local moments are suitable
(e.g., FM layered manganites~\cite{Dessau_1998}), and the orbital and momentum dependence of the self-energy must be disentangled carefully.

\paragraph{Acknowledgements.}
We acknowledge the financial support by Grant-in-Aids for Scientific Research (JSPS KAKENHI) [Grant No. 20K14423 (YN), 21H01041 (YN), 19H02594 (SS), and 19H05825 (RA)] and MEXT as ``Program for Promoting Researches on the Supercomputer Fugaku'' (Basic Science for Emergence and Functionality in Quantum Matter ---Innovative Strongly-Correlated Electron Science by Integration of ``Fugaku'' and Frontier Experiments---) (Project ID: JPMXP1020200104).

\bibliography{main}

\phantom{\cite{Sakai_2016,Yamaji_2011}}

\clearpage

\appendix

\setcounter{section}{0}
\setcounter{equation}{0}
\setcounter{figure}{0}
\setcounter{table}{0}
\setcounter{page}{1}
\renewcommand{\theequation}{S\arabic{equation}}
\renewcommand{\thefigure}{S\arabic{figure}}
\renewcommand{\thetable}{S\arabic{table}}
\renewcommand{\thesection}{S\arabic{section}}

\begin{center}
    {\bf Supplemental Materials for \\ ``Fermi surface expansion above critical temperature in a Hund ferromagnet'' \\ } 
    \vspace{0.2cm}
    {\bf \small Yusuke Nomura, Shiro Sakai, and Ryotaro Arita}
 \end{center}


\subsection{Supplemental data for single-orbital model}

\begin{figure}[H]
\vspace{2cm}
\begin{center}
\includegraphics[width=0.48\textwidth]{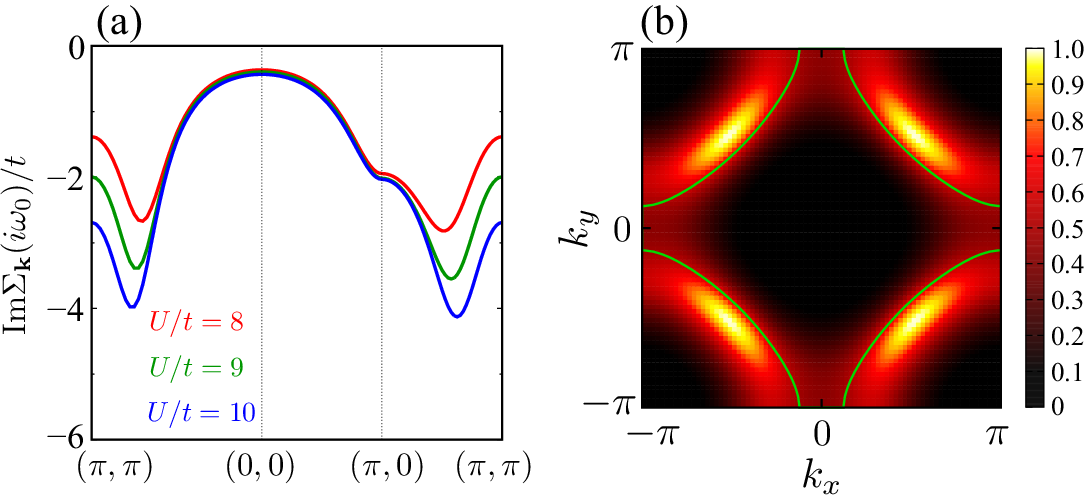}
\caption{
A single-orbital-model counterpart of Fig. 2 in the main text (correlation effects in the AFM fluctuating regime).
Results for $n=0.95$ (5 \% hole doping) and $T=0.06$ for the single-orbital model with $t'=-0.2$. 
(a) Imaginary part of the periodized self-energy for $U=8$, 9, and 10,  
and (b) FS [$-\beta G_{\bf k} (\tau \! = \! \beta/2)$ normalized by its maximum value] at $U=9$. 
The panel (a) employs the common $y$-axis scale as that in Fig. 2(a).
The green curve in (b) shows the FS at $U=0$. 
}
\label{Fig_1orb_AFM}
\end{center}
\end{figure}

As a comparative study, we have performed cDMFT calculations for the single-orbital Hubbard model on the square lattice.
Figure \ref{Fig_1orb_AFM} shows a single-orbital-model counterpart of Fig.~2 in the main text.
The calculation was done for $n=0.95$ (in the AFM-fluctuation regime), which is the same doping ratio (5 \% hole doping) as that in Fig.~2. The results show a qualitatively similar feature to those in the two-orbital case in Fig.~2, albeit with larger interaction values. 

\begin{figure}[H]
\vspace{0cm}
\begin{center}
\includegraphics[width=0.48\textwidth]{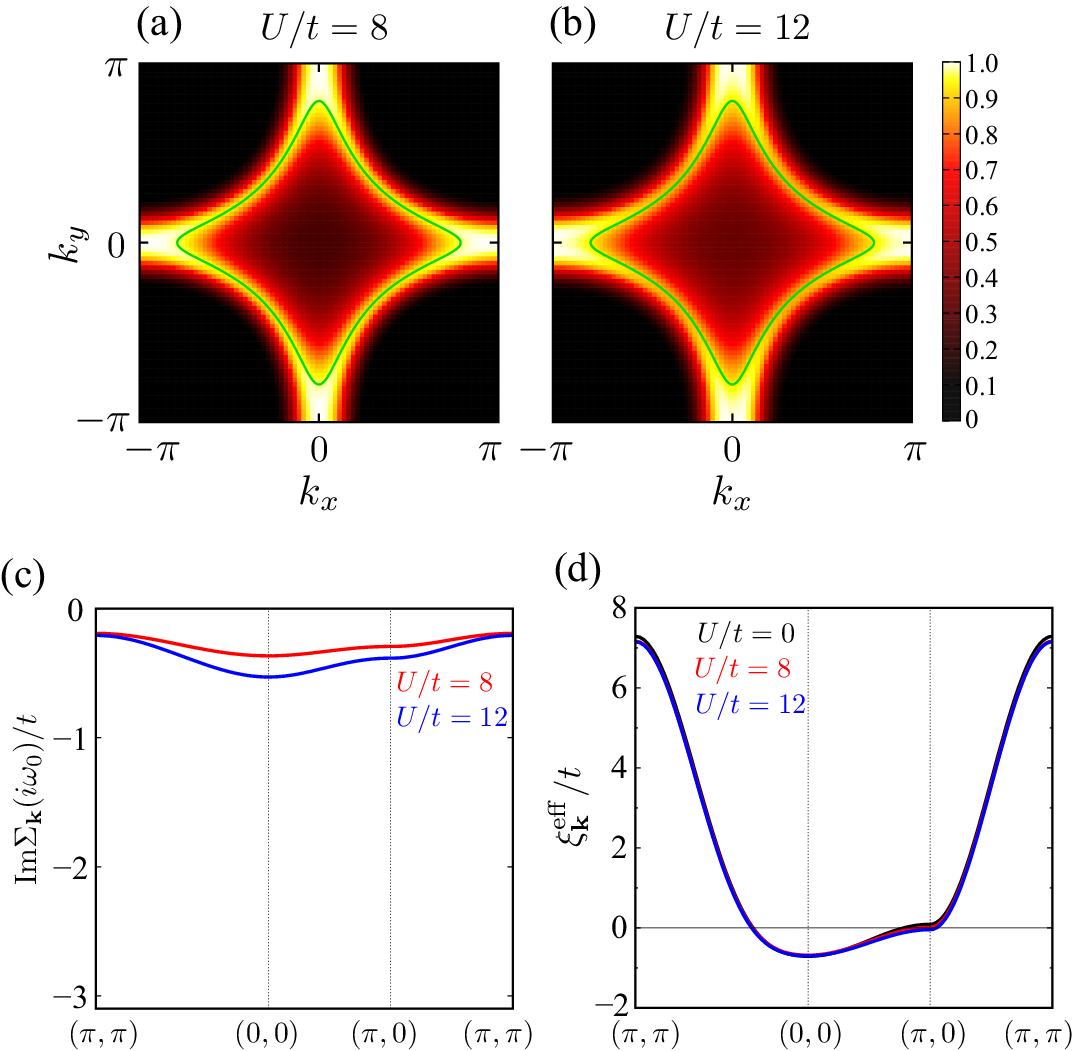}
\caption{
The cDMFT results for the FM-fluctuation regime [$n=0.5$ (50 \% hole doping) and $T=0.06$] in the single-orbital Hubbard model with $t'=-0.4$.
Here, we use a larger $|t'|$ value compared to that employed in the main text ($t'=-0.2$) to stabilize the FM spin correlation in the single-orbital model. 
(a,b) The FS [$-\beta G_{\bf k} (\tau \! = \! \beta/2)$ normalized by its maximum value], 
(c) the imaginary part of the periodized self-energy, 
and 
(d) the effective dispersion $\xi_{\bf k}^{\rm eff} = \epsilon_{\bf k} + {\rm Re }\Sigma_{\bf k}(0) - \mu $ [where ${\rm Re }\Sigma_{\bf k}(0)$ is approximated by ${\rm Re }\Sigma_{\bf k}( i \omega_0)$]. 
The green curves in (a,b) show the FS at $U=0$.
The panel (c) employs the common $y$-axis scale as that in Fig. 3(c).
}
\label{Fig_1orb_FM}
\end{center}
\end{figure}

Figure~\ref{Fig_1orb_FM} shows the results for the FM fluctuating regime in the single-orbital model with the FM fluctuation being stabilized by increasing $|t'|$ to $|t'|= 0.4$. 
As Figs.~\ref{Fig_1orb_FM}(c) and ~\ref{Fig_1orb_FM}(d) show, 
the self-energy effect is much weaker than that in the two-orbital model [Fig.~3], and we cannot see a clear deviation of the FS (aside from the broadening by temperature and weakly momentum-dependent imaginary part of the self-energy) from the $U=0$ one [Figs.~\ref{Fig_1orb_FM}(a) and ~\ref{Fig_1orb_FM}(b)].

\subsection{Supplemental data for rotationally invariant model}

\begin{figure}[H]
\vspace{0cm}
\begin{center}
\includegraphics[width=0.48\textwidth]{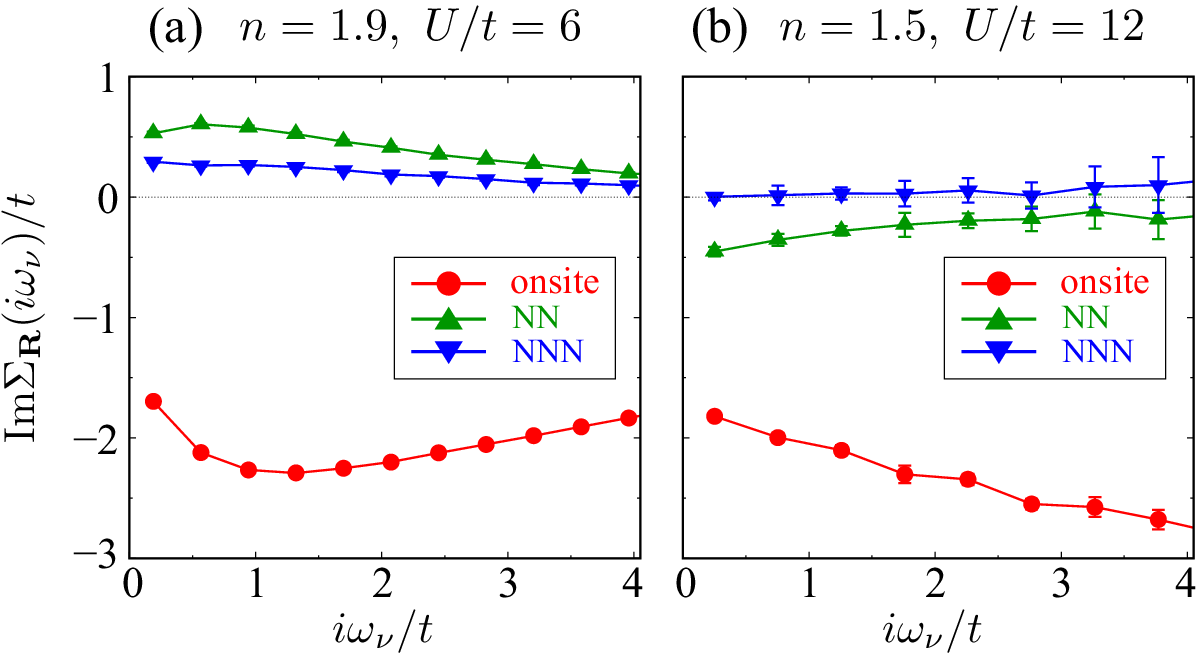}
\caption{
Imaginary part of the real-space self-energy inside the cluster for the two-orbital Hubbard model.
(a) Result for $n=1.9$ and $U=4J=6$. 
(b) Result for $n=1.5$ and $U=4J=12$.}
\label{Fig_Sigma_R}
\end{center}
\end{figure}

Figure~\ref{Fig_Sigma_R} shows the real-space data of the imaginary part of the self-energy. 
When transformed to momentum space, the results in Figs.~\ref{Fig_chi_s_sigma_raw}(c) and \ref{Fig_chi_s_sigma_raw}(d) are obtained.

\begin{figure}[H]
\vspace{0cm}
\begin{center}
\includegraphics[width=0.48\textwidth]{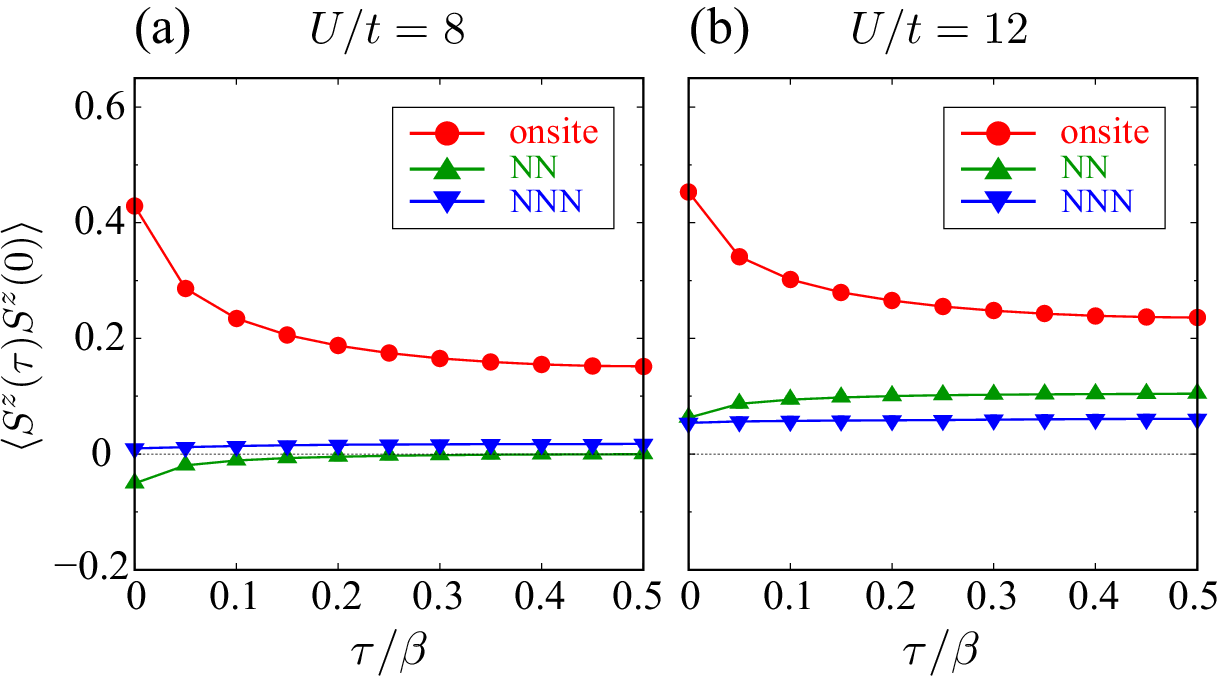}
\caption{
Real-space spin-spin correlation function inside the cluster for the two-orbital Hubbard model at $n=1.5$. 
The panel (b) (result for $U=4J=12$) is the same as Fig.~\ref{Fig_chi_s_sigma_raw}(b) in the main text, reproduced here to compare with the result for $U=4J=8$ in (a). 
}
\label{Fig_n1.5_chi_s}
\end{center}
\end{figure}

Figure~\ref{Fig_n1.5_chi_s} shows the real-space spin-spin correlation function at $n=1.5$. 
Whereas the nonlocal correlations are small and weakly antiferromagnetic at $U=8$ [Fig.~\ref{Fig_n1.5_chi_s}(a)], by increasing the interaction strength, a ferromagnetic spin correlation appears [Fig.~\ref{Fig_n1.5_chi_s}(b)].

\subsection{Supplemental data for density-density-type interaction case}

\begin{figure}[t]
\vspace{0.0cm}
\begin{center}
\includegraphics[width=0.48\textwidth]{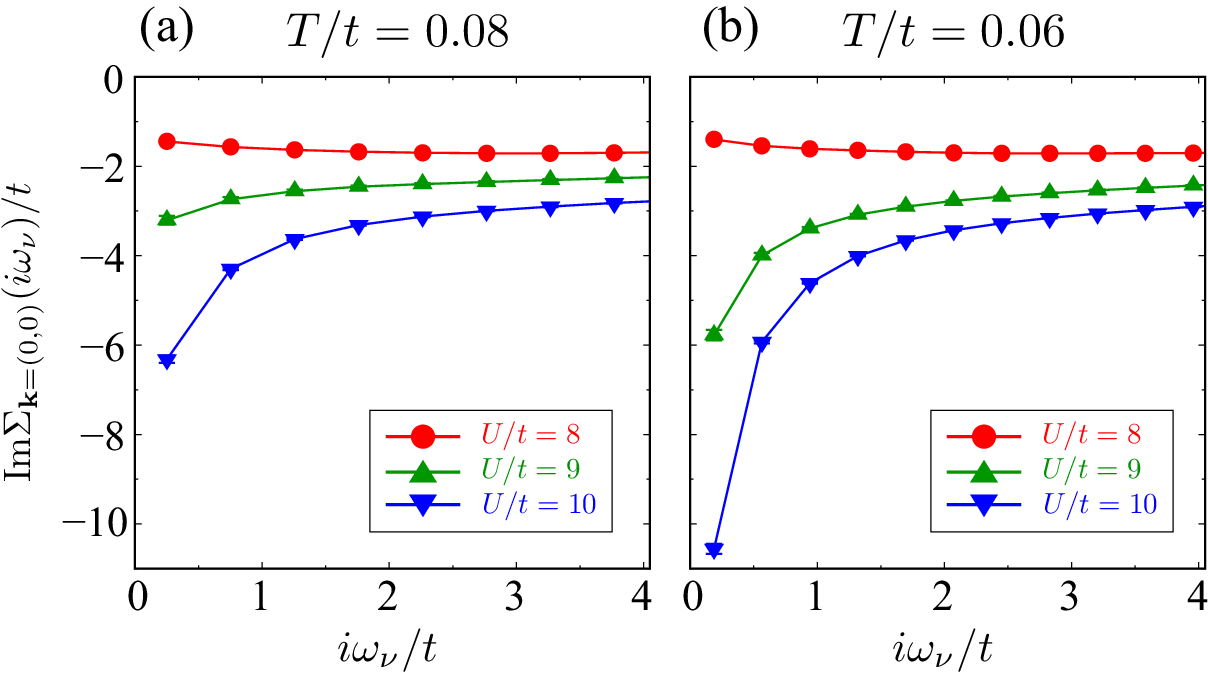}
\caption{
Imaginary part of the self-energy at ${\bf k}=(0,0)$ for the two-orbital model with the density-density-type interactions at $n=1.5$.
The panel (a) employs the same temperature ($T=0.08$) as that for the rotationally invariant case in the main text.
The penal (b) shows the results for a lower temperature ($T=0.06$).
}
\label{Fig_Ising_n1.5_Sigma_00}
\end{center}
\end{figure}

Here, we show the results for the two-orbital model with only density-density-type interactions. 
Figure \ref{Fig_Ising_n1.5_Sigma_00} shows the imaginary part of the self-energy at ${\bf k}=(0,0)$. 
At the same temperature $T=0.08$, the self-energy is more divergent even at a smaller $U$ ($U=10$) than the $U=12$ result for the rotationally invariant case  [Fig.~\ref{Fig_chi_s_sigma_raw}(d)].
As the temperature decreases, the divergence of the self-energy at $U=9$ and $U=10$ becomes more prominent. 

\begin{figure}[t]
\vspace{0cm}
\begin{center}
\includegraphics[width=0.48\textwidth]{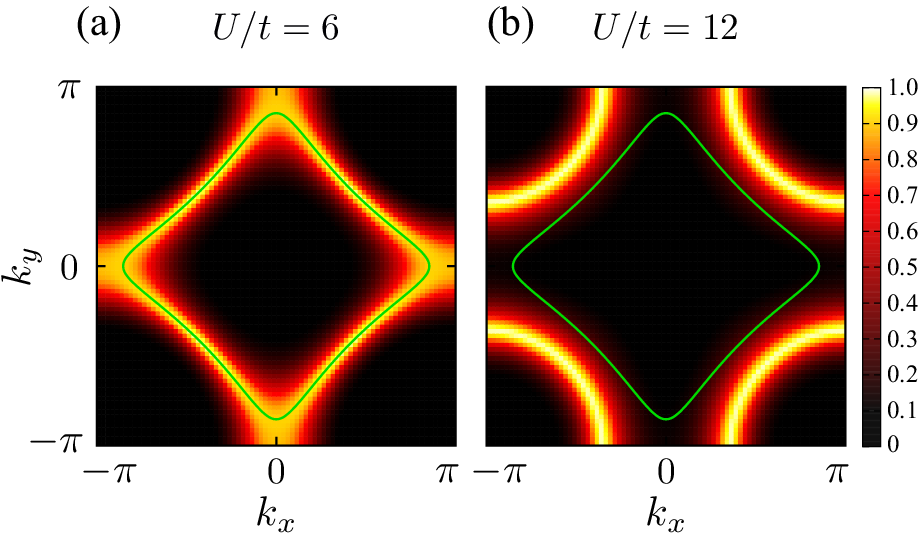}
\caption{
FSs [$-\beta G_{\bf k} (\tau \! = \! \beta/2)$ normalized by its maximum value] at (a) $U=4J=6$ and (b) $U=4J=12$ for the two-orbital model with the density-density-type interactions at $n=1.5$ and $T=0.06$. 
The green curves indicate the FS in the noninteracting case. 
}
\label{Fig_Ising_n1.5_T0.06_FS}
\end{center}
\end{figure}

Figure~\ref{Fig_Ising_n1.5_T0.06_FS} shows the evolution of the FS for the density-density-type interaction.
At $U=12$, we see a change of the FS qualitatively similar to that
in the rotationally invariant case [Fig.~\ref{Fig_n1.5_FS_Sigma}(b)].
But the change is clearer in the density-density-type interaction case, in accord with a stronger divergence in the self-energy. 

\begin{figure}[H]
\vspace{0.0cm}
\begin{center}
\includegraphics[width=0.3\textwidth]{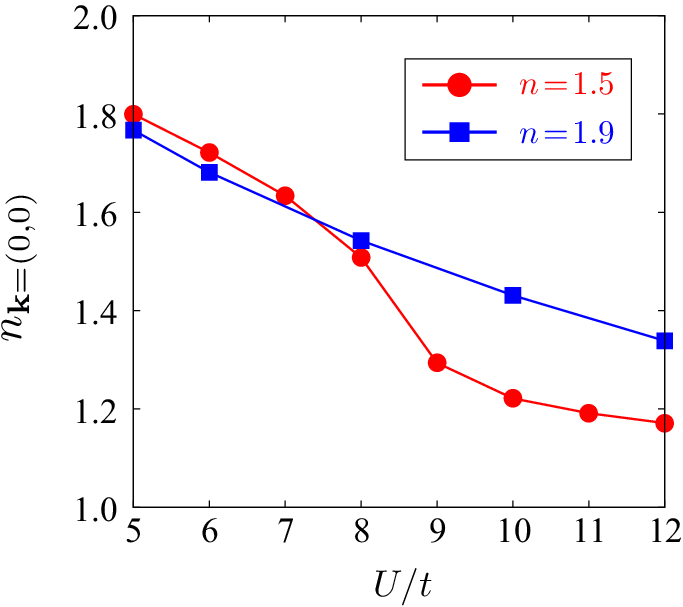}
\caption{
Momentum-space electron occupation per orbital at ${\bf k} = (0,0)$ for the two-orbital model with the density-density-type interactions. 
The results for $n=1.9$ and $n=1.5$ are compared at $T=0.06$. 
}
\label{Fig_Ising_Delta_Nk_Gamma}
\end{center}
\end{figure}

Figure~\ref{Fig_Ising_Delta_Nk_Gamma} shows a comparison of the electron occupation at ${\bf k}=(0,0)$ at $T=0.06$ between $n=1.9$ and $n=1.5$. 
In both cases, the occupation decreases as the interaction increases.
This is natural because, in the limit of a large interaction, the electron occupation in momentum space will become flat. 
However, in the case of $n=1.5$, on top of this effect, the effective ``Hund's coupling'', which competes with the effective ``crystal-field splitting'', plays a role in the electron redistribution in momentum space 
(note that at $n=1.9$, the spin correlation is antiferromagnetic, and the sign of ``Hund's coupling'' is negative).  
When the self-energy becomes divergent for $U\gtrsim9$ [see Fig.~\ref{Fig_Ising_n1.5_Sigma_00}(b)], we clearly see that the occupation at ${\bf k}=(0,0)$ is lowered compared to the $n=1.9$ case.  
This indicates that the effective ``Hund's coupling'' surpasses the effective ``crystal-field splitting'', leading to a formation of a ``high-spin''-like state in momentum space.

\begin{figure}[H]
\vspace{0.0cm}
\begin{center}
\includegraphics[width=0.42\textwidth]{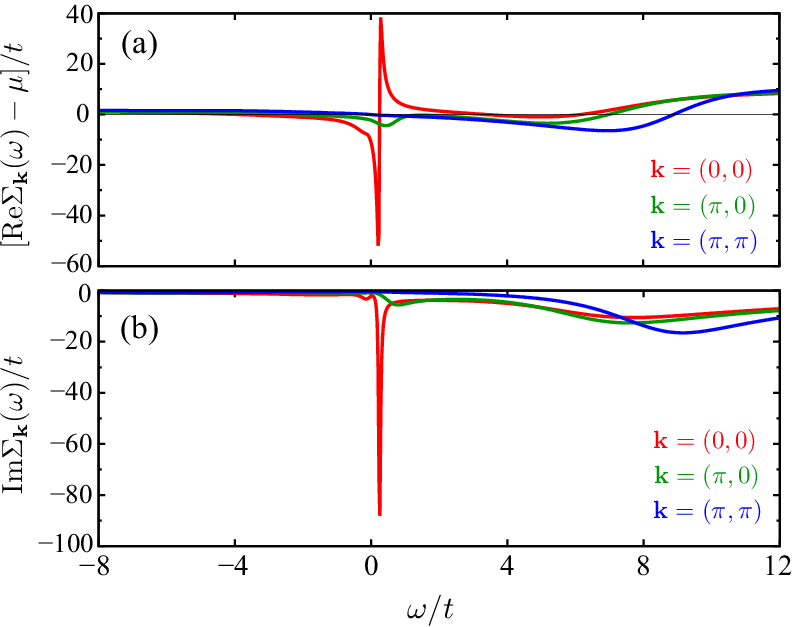}
\caption{
(a) Real and (b) imaginary parts of the self-energy $\Sigma_{\bf k}(\omega)$ along the real frequency axis at $n=1.5$, $U=4J=12$, and $T=0.06$ for the density-density-type interaction model [same parameters as those employed in Fig. 4(b)]. 
}
\label{Fig_Ising_Sigma_real_freq}
\end{center}
\end{figure}

Figure~\ref{Fig_Ising_Sigma_real_freq} shows the self-energy along the real frequency axis in the regime of momentum-space ``high-spin'' formation ($U=12$, $T=0.06$ with the density-density-type interaction), i.e., strong-coupling FM-fluctuation regime. 
We see a sharp peak in the self-energy at ${\bf k}=(0,0)$, in accord with the significant electronic structure modulation [Fig.~4(b)] and the redistribution of the momentum-space electron occupation (Fig.~\ref{Fig_Ising_Delta_Nk_Gamma}). 
This singularly sharp peak in the self-energy indicates a presence of a hidden fermionic excitation emergent from electron correlations~\cite{Sakai_2016}. 
In fact, there seem to be two different types of low-energy excitations in the present two-orbital system: 
Quasiparticles propagating on the lattice and localized electrons forming the local moment through Hund's coupling. 
This is analogous to the situation of doped Mott insulators in the single-orbital Hubbard model, where the localized electrons emerge from a trapping by doped holes~\cite{Yamaji_2011}. 
The propagating electrons sometimes transit to the localized state, which generates a singular self-energy in its frequency dependence.

\end{document}